\documentclass[preprintnumbers,amsmath,amssymbm,preprint]{revtex4}
\usepackage{epsfig}
\usepackage{graphicx}
\usepackage{amssymb}

\begin{document}
\title{Further evidence for the non-existence of a unified hoop conjecture}
\author{Shahar Hod}
\affiliation{The Ruppin Academic Center, Emeq Hefer 40250, Israel}
\affiliation{ } \affiliation{The Hadassah Institute, Jerusalem
91010, Israel}
\date{\today}

\begin{abstract}
\ \ \ The hoop conjecture, introduced by Thorne almost five decades
ago, asserts that black holes are characterized by the
mass-to-circumference relation $4\pi {\cal M}/{\cal C}\geq1$, whereas
horizonless compact objects are characterized by the opposite
inequality $4\pi {\cal M}/{\cal C}<1$ (here ${\cal C}$ is the
circumference of the smallest ring that can engulf the
self-gravitating compact object in all azimuthal directions).
It has recently been proved that a necessary condition for the
validity of this conjecture in horizonless spacetimes of spatially
regular charged compact objects is that the mass ${\cal M}$ be
interpreted as the mass contained within the engulfing sphere (and
not as the asymptotically measured total ADM mass). In the present
paper we raise the following physically intriguing question: Is it possible to
formulate a {\it unified} version of the hoop conjecture which is
valid for both black holes and horizonless compact objects? In order
to address this important question, we analyze the
behavior of the mass-to-circumference ratio of Kerr-Newman black
holes. We explicitly prove that if the mass ${\cal M}$ in the hoop
relation is interpreted as the quasilocal Einstein-Landau-Lifshitz-Papapetrou and Weinberg mass contained within the black-hole
horizon, then these charged and spinning black holes are
characterized by the {\it sub}-critical mass-to-circumference ratio
$4\pi {\cal M}/{\cal C}<1$. Our results provide evidence for the
non-existence of a unified version of the hoop conjecture which is
valid for both black-hole spacetimes and spatially regular
horizonless compact objects.
\end{abstract}
\bigskip
\maketitle

\section{Introduction}

The influential hoop conjecture has been suggested by Thorne
\cite{Thorne} as a simple necessary and sufficient condition for the
formation of black holes in dynamical gravitational collapse
scenarios. In particular, the hoop criterion asserts that a
self-gravitating matter configuration of mass ${\cal M}$ would
collapse to form a black hole if and only if a circular hoop of a
critical circumference ${\cal C}_{\text{critical}}=4\pi {\cal M}$
can be placed around the self-gravitating matter distribution and
rotated in $360^{\circ}$ to form an engulfing sphere. The hoop
conjecture therefore implies the simple relation \cite{Thorne}
\begin{equation}\label{Eq1}
{\cal H}\equiv{{4\pi {\cal M}}\over{{\cal C}}}\geq1\ \
\Longleftrightarrow\ \ \text{Black-hole horizon exists}\  .
\end{equation}

In his original formulation of the hoop conjecture, Thorne
\cite{Thorne} has not provided an explicit definition for the mass
term ${\cal M}$ in his dimensionless mass-to-circumference ratio
${\cal H}\equiv 4\pi {\cal M}/{\cal C}$ \cite{Noteunit}.
Interestingly, as explicitly demonstrated in \cite{Leo,Bon,Andre},
the hoop conjecture (\ref{Eq1}) can be violated in curved spacetimes
of horizonless charged compact objects if ${\cal M}$ is interpreted
as the total ADM mass of the spacetime. In particular, as nicely
shown in \cite{Andre}, spherically symmetric horizonless thin shells
of radius $R$, electric charge $Q$, and total ADM mass $M$ can be
characterized by the dimensionless relation $M/R\to 1^-$ in the
$Q/M\to1^-$ limit. Thus, these horizonless charged shells are
characterized by the super-critical dimensionless relation $4\pi
M/{\cal C}\to2^-$ and can therefore violate the hoop conjecture
(\ref{Eq1}) if ${\cal M}$ is interpreted as the total ADM mass $M$
of the spacetime.

Intriguingly, it has recently been proved \cite{Hodst} (see also
\cite{Peng}) that if the mass ${\cal M}$ is interpreted as the
gravitating mass $M_{\text{in}}\equiv M(R)$ contained within an
engulfing sphere of radius $R$ \cite{Notedeg} (and not as the total
ADM mass of the spacetime), then horizonless self-gravitating
charged compact objects are characterized by the sub-critical
dimensionless relation ${\cal H}<1$ and therefore respect the hoop
conjecture (\ref{Eq1}), see also the physically interesting related works \cite{New1,New2,New3,New4,New5}.

The interesting physical results presented in
\cite{Leo,Bon,Andre,Hodst,Peng} indicate that if there is any hope
to formulate the hoop conjecture (\ref{Eq1}) in a {\it unified} way,
which is valid for {\it both} black-hole spacetimes and spatially
regular (horizonless) self-gravitating compact objects, then the
mass term ${\cal M}$ should not be interpreted as the total ADM mass
of the spacetime. In particular, {\it if} a valid and unified
formulation of the conjecture (\ref{Eq1}) do exists, then the
results presented in \cite{Leo,Bon,Andre,Hodst,Peng} indicate that
the mass ${\cal M}$ should be interpreted as the mass
$M_{\text{in}}$ contained within the boundaries of the compact
object (a black hole or a self-gravitating horizonless object).

Motivated by the intriguing results of
\cite{Leo,Bon,Andre,Hodst,Peng}, in the present compact paper we
raise the following physically interesting question: Is it possible
to formulate the hoop conjecture (\ref{Eq1}) in a {\it unified} way
which is valid for both black-hole spacetimes and spatially regular
horizonless compact objects?

In order to address this intriguing question, we shall analyze the
functional behavior of the dimensionless mass-to-circumference ratio
${\cal H}(Q,a)\equiv 4\pi {\cal M}/{\cal C}$ of charged and spinning
Kerr-Newman black-hole spacetimes (here $Q$ and $a$ are respectively
the electric charge and the angular momentum per unit mass of the
Kerr-Newman black holes). Interestingly, we shall explicitly show
below that if the mass ${\cal M}$ in the hoop relation (\ref{Eq1})
is interpreted as the quasilocal Einstein-Landau-Lifshitz-Papapetrou and Weinberg mass $M_{\text{in}}$ contained within the
black-hole horizon, then charged and spinning Kerr-Newman black
holes are characterized by the sub-critical mass-to-circumference
ratio $4\pi M_{\text{in}}/{\cal C}<1$ and therefore violate the hoop
conjecture (\ref{Eq1}).

\section{The hoop conjecture in charged and spinning Kerr-Newman black-hole spacetimes}

A Kerr-Newman black-hole spacetime of asymptotically measured ADM
mass $M$, electric charge $Q$ \cite{Noteqp}, and angular momentum
$J=Ma$ is described by the curved line element
\cite{Bar,Chan,Shap,Notebl}
\begin{eqnarray}\label{Eq2}
ds^2=-{{\Delta-a^2\sin^2\theta}\over{\rho^2}}dt^2+{{\rho^2}\over{\Delta}}dr^2
-{{2a\sin^2\theta(2Mr-Q^2)}\over{\rho^2}}dt d\phi
+\rho^2 d\theta^2\nonumber \\
+{{(r^2+a^2)^2-a^2\Delta\sin^2\theta}\over{\rho^2}}\sin^2\theta d\phi^2\ ,
\end{eqnarray}
where
\begin{equation}\label{Eq3}
\Delta\equiv r^2-2Mr+Q^2+a^2\ \ \ \ ; \ \ \ \ \rho^2\equiv
r^2+a^2\cos^2\theta\  .
\end{equation}
The radii
\begin{equation}\label{Eq4}
r_{\pm}=M\pm (M^2-Q^2-a^2)^{1/2}\
\end{equation}
of the (outer and inner) black-hole horizons are determined by the
zeros of the metric function $\Delta(r)$.

Substituting $dt=dr=d\theta=0$, $\theta=\pi/2$, and
$\Delta\phi=2\pi$ into the curved line element (\ref{Eq2}), one
finds the simple functional expression
\begin{equation}\label{Eq5}
{\cal
C}_{\text{eq}}=2\pi{{r^2_++a^2}\over{r_+}}\cdot[1+O(\epsilon)]\
\end{equation}
for the equatorial circumference of an engulfing ring which is
located at a fixed radial coordinate $r=r_+\cdot[1+O(\epsilon)]$
just outside ($\epsilon\ll1$) the outer horizon (\ref{Eq4}) of the
Kerr-Newman black-hole \cite{Noteeqpo}. Taking cognizance of Eq.
(\ref{Eq4}), one can express the black-hole equatorial circumference
(\ref{Eq5}) in the dimensionless form \cite{Noteind}
\begin{equation}\label{Eq6}
{\bar{\cal C}_{\text{eq}}}({\bar Q},{\bar
a})=4\pi\cdot \Bigg[1-{{{\bar Q}^2}\over{2\big(1+\sqrt{1-{\bar Q}^2-{\bar a}^2}\big)}}\Bigg]\  ,
\end{equation}
where
\begin{equation}\label{Eq7}
{\bar{\cal C}}\equiv {{{\cal C}}\over{M}}\ \ \ \ ; \ \ \ \ {\bar
Q}\equiv {{Q}\over{M}}\ \ \ \ ; \ \ \ \ {\bar a}\equiv
{{a}\over{M}}\
\end{equation}
are respectively the dimensionless circumference, the dimensionless
electric charge, and the dimensionless angular momentum of the
Kerr-Newman black hole.

Before proceeding, it is worth pointing out that the hoop relation
(\ref{Eq1}) is respected by all Kerr-Newman black holes if the mass term
${\cal M}$ is interpreted as the {\it total} ADM mass $M$ of the
spacetime. In particular, taking cognizance of Eqs. (\ref{Eq6}) and
(\ref{Eq7}), one finds the simple dimensionless relation
\begin{equation}\label{Eq8}
{\cal H}({\bar Q},{\bar a})\equiv {{4\pi M_{\text{tot}}}\over{{\cal
C}}}= \Bigg[1-{{{\bar Q}^2}\over{2\big(1+\sqrt{1-{\bar Q}^2-{\bar
a}^2}\big)}}\Bigg]^{-1}\geq1\
\end{equation}
for the charged and spinning Kerr-Newman black holes.

However, as emphasized above, it has been demonstrated explicitly in
\cite{Leo,Bon,Andre} that the hoop conjecture (\ref{Eq1}) can be
violated by spatially regular horizonless charged objects if the
mass ${\cal M}$ is interpreted as the total ADM mass of the
spacetime. This fact implies that a unified version of the hoop
conjecture, which should be valid for both black-hole spacetimes and
horizonless compact objects, cannot be formulated in terms of the
total ADM mass of the spacetime.

On the other hand, it has been explicitly proved in
\cite{Hodst,Peng} that the hoop relation (\ref{Eq1}) is respected by
horizonless charged compact objects if the mass term ${\cal M}$ is
interpreted as the mass $M_{\text{in}}(R)$ contained within the
engulfing sphere of radius $R$. This observation indicates that {\it
if} a unified formulation of the hoop conjecture (\ref{Eq1}) exists,
then the mass ${\cal M}$ should be interpreted as the mass contained
within the engulfing sphere \cite{Notedeg}.

We shall therefore test the validity of the hoop relation
(\ref{Eq1}) for Kerr-Newman black holes with the mass term ${\cal
M}$ interpreted as the mass $M_{\text{in}}(r_+)$ contained within
the black-hole horizon. To this end, we first note that, in a
physically interesting paper, Aguirregabiria, Chamorro, and
Virbhadra \cite{Agu} have analyzed the energy distribution in the
Kerr-Newman black-hole spacetime using various physical
prescriptions (namely, the Einstein, the Landau-Lifshitz, the
Papapetrou, and the Weinberg prescriptions \cite{Agu}) for calculating energy
densities in general relativity. Interestingly, it has been explicitly shown
in \cite{Agu} that the Einstein, Landau-Lifshitz, Papapetrou, and Weinberg prescriptions provide the same compact
analytical expression \cite{Agu,NoteRN}
\begin{equation}\label{Eq9}
M_{\text{in}}\equiv
M_{\text{in}}(r=r_+)=M-{{Q^2}\over{4r_+}}\Big[1+{{r^2_++a^2}\over{ar_+}}\cdot\arctan\Big({{a}\over{r_+}}\Big)\Big]\
\end{equation}
for the quasilocal Einstein-Landau-Lifshitz-Papapetrou and Weinberg mass $M_{\text{in}}$ contained within the horizon of a
Kerr-Newman black hole. Taking cognizance of Eqs. (\ref{Eq4}) and
(\ref{Eq9}), one can express the engulfed gravitational mass
$M_{\text{in}}$ of the charged and spinning Kerr-Newman black hole
(\ref{Eq2}) explicitly in terms of its dimensionless physical
parameters ${\bar Q}$ and ${\bar a}$:
\begin{equation}\label{Eq10}
{\bar M}_{\text{in}}({\bar Q},{\bar a})=1-{{{\bar
Q}^2}\over{4\big(1+\sqrt{1-{\bar Q}^2-{\bar
a}^2}\big)}}\Bigg[1+{{\big(1+\sqrt{1-{\bar Q}^2-{\bar
a}^2}\big)^2+{\bar a}^2}\over{{\bar a}\big(1+\sqrt{1-{\bar
Q}^2-{\bar a}^2}\big)}}\cdot\arctan\Big({{{\bar
a}}\over{1+\sqrt{1-{\bar Q}^2-{\bar a}^2}}}\Big)\Bigg]\  ,
\end{equation}
where
\begin{equation}\label{Eq11}
{\bar M}_{\text{in}}\equiv {{M_{\text{in}}}\over{M}}\  .
\end{equation}

In Table \ref{Table1} we present the dimensionless
mass-to-circumference ratio [see Eqs. (\ref{Eq6}) and (\ref{Eq10})]
\begin{equation}\label{Eq12}
{\cal H}({\bar Q},{\bar a})\equiv {{4\pi{\cal M}}\over{{\cal
C}_{\text{eq}}}}\ \ \ \ ; \ \ \ \ {\cal M}=M_{\text{in}}(r_+)\
\end{equation}
of charged and spinning Kerr-Newman black holes for various values
of the black-hole dimensionless physical parameters ${\bar Q}$ and
${\bar a}$. Intriguingly, the data presented in Table \ref{Table1}
reveals the fact that all charged and spinning (${\bar Q}\neq0,{\bar
a}\neq0$) Kerr-Newman black holes are characterized by the {\it
sub}-critical mass-to-circumference relation \cite{Notespb}
\begin{equation}\label{Eq13}
{\cal H}({\bar Q},{\bar a})<1\  .
\end{equation}

\begin{table}[htbp]
\centering
\begin{tabular}{|c|c|c|c|c|c|c|c|c|c|}
\hline ${\cal H}({\bar Q},{\bar a})$ & \ ${\bar a}=0.0$\ \ & \
${\bar a}=0.2$\ \ & \
${\bar a}=0.4$\ \ & \ ${\bar a}=0.6$\ \ & \ ${\bar a}=0.8$\ \ & \ ${\bar a}=1.0$\ \  \\
\hline \ ${\bar Q}=0.0$\ \ \ &\ \ 1.0000\ \ \ &\ \ 1.0000\
\ \ &\ \ 1.0000\ \ \ &\ \ 1.0000\ \ \ &\ \ 1.0000\ \ \ &\ \ 1.0000 \ \ \\
\hline \ ${\bar Q}=0.2$\ \ \ &\ \ 1.0000\ \ \ &\ \ 0.9999\
\ \ &\ \ 0.9998\ \ \ &\ \ 0.9996\ \ \ &\ \ 0.9989\ \ \ &\ \ $---$ \ \ \\
\hline \ ${\bar Q}=0.4$\ \ \ &\ \ 1.0000\ \ \ &\ \ 0.9998\
\ \ &\ \ 0.9993\ \ \ &\ \ 0.9980\ \ \ &\ \ 0.9944\ \ \ &\ \ $---$ \ \ \\
\hline \ ${\bar Q}=0.6$\ \ \ &\ \ 1.0000\ \ \ &\ \ 0.9995\
\ \ &\ \ 0.9978\ \ \ &\ \ 0.9934\ \ \ &\ \ 0.9579\ \ \ &\ \ $---$ \ \ \\
\hline \ ${\bar Q}=0.8$\ \ \ &\ \ 1.0000\ \ \ &\ \ 0.9986\
\ \ &\ \ 0.9929\ \ \ &\ \ 0.9471\ \ \ &\ \ $---$\ \ \ &\ \ $---$ \ \ \\
\hline \ ${\bar Q}=1.0$\ \ \ &\ \ 1.0000\ \ \ &\ \ $---$\
\ \ &\ \ $---$\ \ \ &\ \ $---$\ \ \ &\ \ $---$\ \ \ &\ \ $---$ \ \ \\
\hline
\end{tabular}
\caption{The dimensionless mass-to-circumference ratio ${\cal
H}({\bar Q},{\bar a})\equiv 4\pi{\cal M}/{\cal C}_{\text{eq}}$ of
charged and spinning Kerr-Newman black holes. Here ${\cal M}$ is
interpreted as the quasilocal Einstein-Landau-Lifshitz-Papapetrou and Weinberg mass contained within the black-hole horizon and
${\cal C}_{\text{eq}}$ is the equatorial circumference along the
black-hole outer horizon [see Eqs. (\ref{Eq6}) and (\ref{Eq10})].
Interestingly, the dimensionless ratio ${\cal H}({\bar Q},{\bar a})$
is found to be a monotonically decreasing function of the black-hole
physical parameters ${\bar a}$ and ${\bar Q}$. In particular, one
finds that charged and spinning Kerr-Newman black holes are
characterized by the {\it sub}-critical mass-to-circumference ratio
${\cal H}({\bar Q}\neq0,{\bar a}\neq0)\equiv 4\pi{\cal M}/{\cal
C}_{\text{eq}}<1$.} \label{Table1}
\end{table}

In particular, from the data presented in Table \ref{Table1} one
finds that, for a given value of the black-hole electric charge
${\bar Q}$, the dimensionless ratio ${\cal H}$ is a monotonically
decreasing function of the black-hole angular momentum parameter
${\bar a}$. In addition, one finds that, for a given value of the
black-hole angular momentum ${\bar a}$, the mass-to-circumference
ratio ${\cal H}$ is a monotonically decreasing function of the
black-hole electric charge ${\bar Q}$. These facts indicate that the
dimensionless mass-to-circumference function ${\cal H}({\bar
Q},{\bar a})$ is minimized by an extremal black hole. Substituting
${\bar Q}^2+ {\bar a}^2=1$ for extremal Kerr-Newman black holes
\cite{Noteext} into Eqs. (\ref{Eq6}) and (\ref{Eq10}), one obtains
the compact functional expression [see Eq. (\ref{Eq12})]
\begin{equation}\label{Eq14}
{\cal H}^{\text{ext}}({\bar a})={{1-{{1-{\bar
a}^2}\over{4}}\big[1+{{1+{\bar a}^2}\over{{\bar
a}}}\cdot\arctan({\bar a})\big]}\over{1-{{1-{\bar a}^2}\over{2}}}}\
\end{equation}
for the mass-to-circumference ratio of extremal Kerr-Newman black
holes. From (\ref{Eq14}) one finds
\begin{equation}\label{Eq15}
\text{min}_{{\bar Q},{\bar a}}\{{\cal H}\}=0.9468\ \ \ \ \text{for}\
\ \ \ ({\bar Q},{\bar a})=(0.7776,0.6287)\  .
\end{equation}

\section{Summary and Discussion}

The Thorne hoop conjecture \cite{Thorne} serves as a boundary
between black-hole configurations and horizonless compact objects.
In particular, this physically influential conjecture asserts that
black holes should be characterized by the mass-to-circumference inequality
$4\pi {\cal M}/{\cal C}\geq1$, whereas horizonless compact objects
should be characterized by the opposite inequality $4\pi {\cal
M}/{\cal C}<1$.

The physical meaning of the mass term ${\cal M}$ in the hoop
conjecture (\ref{Eq1}) has not been specified by Thorne
\cite{Thorne}. However, it is known that the hoop relation
(\ref{Eq1}) can be violated by self-gravitating horizonless charged
objects if the mass ${\cal M}$ is interpreted as the total ADM mass
of the spacetime \cite{Leo,Bon,Andre}. On the other hand, it has
recently been proved \cite{Hodst,Peng} that spherically symmetric
horizonless charged objects respect the hoop relation (\ref{Eq1}) if
the mass ${\cal M}$ is interpreted as the gravitating mass
$M_{\text{in}}\equiv M(R)$ contained within an engulfing sphere of
radius $R$ (and not as the total mass of the spacetime).

The physical results presented in \cite{Leo,Bon,Andre,Hodst,Peng}
imply that a unified version of the hoop relation (which would be
valid for {\it both} black-hole spacetimes and spatially regular
horizonless compact objects), if it exists, should be formulated in
terms of the mass $M_{\text{in}}(R)$ contained within the boundaries
of the engulfing sphere \cite{Notedeg}.

Motivated by the interesting physical results of
\cite{Leo,Bon,Andre,Hodst,Peng}, in the present compact paper we
have raised the following physically intriguing question: Is it
possible to formulate a {\it unified} version of the hoop conjecture
which is valid for both black-hole spacetimes and spatially regular
horizonless compact objects?

In order to address this physically important question, we have
analyzed the behavior of the dimensionless mass-to-circumference
ratio ${\cal H}({\bar Q},{\bar a})\equiv 4\pi {\cal M}/{\cal C}$ of
charged and spinning Kerr-Newman black holes, where ${\cal
M}=M_{\text{in}}(r_+)$ is the quasilocal Einstein-Landau-Lifshitz-Papapetrou and Weinberg mass contained within the black-hole
horizon. The main results derived in this paper and their physical
implications are as follows:
\newline
(1) While it is well known that neutral spinning Kerr black holes
and spherically symmetric charged Reissner-Nordstr\"om (RN) black
holes saturate the hoop relation [see Eqs. (\ref{Eq4}), (\ref{Eq6}),
and (\ref{Eq10}) with ${\bar Q}=0$ or ${\bar a}=0$]
\begin{equation}\label{Eq16}
{{4\pi M_{\text{in}}}\over{{\cal C}_{\text{eq}}}}=1\ \ \ \ \text{for
Kerr and RN black holes}\  ,
\end{equation}
we have explicitly proved that charged-spinning (${\bar
Q}\neq0,{\bar a}\neq0$) Kerr-Newman black holes are characterized by
the {\it sub}-critical dimensionless mass-to-circumference ratio
(see the data presented in Table \ref{Table1})
\cite{Notemol,Cohen,Xul}
\begin{equation}\label{Eq17}
{{4\pi M_{\text{in}}}\over{{\cal C}_{\text{eq}}}}<1\ \ \ \ \text{for
charged {\it and} spinning Kerr-Newman black holes}\ .
\end{equation}
\newline
(2) It has been shown that, for Kerr-Newman black holes, the
dimensionless mass-to-circumference ratio ${\cal H}({\bar Q},{\bar
a})\equiv{{4\pi M_{\text{in}}}/{{\cal C}_{\text{eq}}}}$ is a
monotonically decreasing function of the black-hole physical
parameters ${\bar Q}$ and ${\bar a}$. In particular, for a given
value ${\bar Q}$ of the black-hole electric charge, the hoop
relation ${\cal H}({\bar a};{\bar Q})$ is minimized by the extremal
Kerr-Newman black hole with ${\bar a}=\sqrt{1-{\bar Q}^2}$.
Likewise, for a given value ${\bar a}$ of the black-hole angular
momentum parameter, the hoop relation ${\cal H}({\bar Q};{\bar a})$
is minimized by the extremal Kerr-Newman black hole with ${\bar
Q}=\sqrt{1-{\bar a}^2}$.
\newline
(3) It has been found that the generic violations of the hoop
conjecture (\ref{Eq1}) with ${\cal M}=M_{\text{in}}(r_+)$ by the
Kerr-Newman family of charged and spinning black holes are always
bounded from below by the relation $0.9468\lesssim {\cal
H}\equiv4\pi M_{\text{in}}/{\cal C}_{\text{eq}}\leq1$ [see Eq.
(\ref{Eq15})] \cite{Notetd}.
\newline
(4) The results presented in this paper provide compelling evidence
for the non-existence of a unified version of the hoop conjecture
which is valid for both black-hole spacetimes and spatially regular
horizonless compact objects. In particular, if the mass term ${\cal
M}$ in the hoop conjecture is interpreted as the total ADM mass of
the spacetime then, as explicitly proved in
\cite{Leo,Bon,Andre,Hodst,Peng}, horizonless charged compact objects
can violate the hoop relation (\ref{Eq1}). On the other hand, in the
present paper we have explicitly proved that if the mass term ${\cal
M}$ in the hoop conjecture is interpreted as the quasilocal Einstein-Landau-Lifshitz-Papapetrou and Weinberg mass contained
within the engulfing sphere, then charged and spinning Kerr-Newman
black holes violate the hoop relation (\ref{Eq1}).

\bigskip
\noindent {\bf ACKNOWLEDGMENTS}

This research is supported by the Carmel Science Foundation. I would
like to thank Yael Oren, Arbel M. Ongo, Ayelet B. Lata, and Alona B.
Tea for stimulating discussions.


\begin{thebibliography}{99}
\bibitem{Noteunit} We use natural units in which $G=c=\hbar=1$. In
these units the fundamental ratio ${\cal H}\equiv 4\pi {\cal
M}/{\cal C}$ is dimensionless.
\bibitem{Notedeg} It is worth emphasizing that the engulfing sphere is formed from
the rotation of the engulfing ring by $360^{\circ}$.
\bibitem{Noteqp} We shall assume $Q\geq0$ without loss of generality.
\bibitem{Noteind} It is interesting to note that one finds from Eq. (\ref{Eq6}) that
the dimensionless equatorial circumference ${\bar{\cal
C}_{\text{eq}}}({\bar Q}=0,{\bar a})$ of a neutral Kerr black hole
is universal in the sense that it is independent of its
dimensionless angular momentum ${\bar a}$.
\bibitem{NoteRN} Note that for non-rotating ($a\to0$) black holes,
one finds from (\ref{Eq9}) the simple relation
$M_{\text{in}}=M-Q^2/2r_+$.
\bibitem{Notespb} Note that spherically symmetric black holes are
characterized by the simple relation
$M_{\text{in}}(r_+)={1\over2}r_+$ for the mass contained within the
black-hole horizon. Thus, if the mass term ${\cal M}$ in the hoop
conjecture (\ref{Eq1}) is interpreted as the mass
$M_{\text{in}}(r_+)$ contained within the black-hole horizon (and
not as the total ADM mass of the spacetime), then all spherically
symmetric black holes would trivially saturate the hoop relation
(\ref{Eq1}).
\bibitem{Noteext} Note that extremal Kerr-Newman black holes are characterized by the simple
relation $r_-=r_+$ [see Eq. (\ref{Eq4})].
\bibitem{Notetd} One therefore concludes that, although Kerr-Newman black holes
violate the word of the hoop relation (\ref{Eq1}) if the mass term
${\cal M}$ is interpreted as the mass $M_{\text{in}}(r_+)$ contained
within the black-hole horizon, these stationary black-hole
spacetimes do respect the spirit of the physically important hoop
conjecture.
\end{thebibliography}
\end{document}